\documentclass{ws-p8-50x6-00}
\newcommand{\eqb}{\begin{equation}}
\newcommand{\eqe}{\end{equation}}
\newcommand{\dmb}{\begin{displaymath}}
\newcommand{\dme}{\end{displaymath}}
\newcommand{\pd}{\partial}
\newcommand{\ep}{\varepsilon}
\newcommand{\eab}{\begin{eqnarray}}
\newcommand{\eae}{\end{eqnarray}}

\newcommand{\e}{\mbox{e}}

\begin{document}

\title{Deconfining by Winding}

\author{Ralf Hofmann}

\address{Max-Planck-Institut f\"ur Physik,\\ 
Werner Heisenberg-Institut,\\  
F\"ohringer Ring 6, 80805 M\"unchen\\ 
E-mail: ralfh@mppmu.mpg.de}


\maketitle

\abstracts{A model for the quantum effective 
description of the vacuum structure of 
thermalized SU(3) 
Yang-Mills theory is proposed. The model is based 
on Abelian projection leading to a 
Ginzburg-Landau theory for the magnetic sector. 
The possibility of topologically 
non-trivial, effective monopole fields in the deconfining phase is 
explored. These fields are assumed to be 
Bogomol'nyi-Prasad-Sommerfield saturated solutions along the compact, 
euclidean time dimension. Accordingly, a gauge invariant interaction for 
the monopole fields is constructed.
 Motivated by the corresponding lattice 
results the vacuum dynamics is assumed to
be dominated by the monopole fields. A reasonable value for the critical 
temperature is obtained, and the partial persistence of 
non-perturbative features in the deconfining 
phase of SU(3) Yang-Mills theory, as it is 
measured on the lattice, follows naturally. 
}

\section{Abelian projection, monopole trajectories, and field theoretic
description}

Quantum Chromodynamics (QCD) and SU(N) Yang-Mills (YM) theory are 
asympotically free (regime of large momenta)\footnote{For QCD this statement is only true if 
the number of active quark flavors is does not exceed a critical value set by the number of colors.}, 
and they exhibit a 
growth of the coupling toward the infrared, perturbatively. The latter is a {\sl necessary} 
condition for a successful explanation 
of the fact that 
the non-collective propagation of color charges over large distances 
has not been observed experimentally. 
Due to lattice results this is commonly believed to be caused by 
the pecularities of the gauge boson interactions. 
Therefore, we restrict ourselves to the discussion of 
pure gluodynamics, that is, SU(N) YM theory. 

\noindent A promising attempt was made 
by Mandelstam and 't Hooft in the 
late seventies and early eighties \cite{HM}. 
Roughly, they proposed to view the QCD and YM 
vacuum as a dual superconductor which forces the 
chromo-electric flux between a pair of largely separated 
test  charges into a tube characterized 
by a constant tension $\sigma$. The corresponding potential grows linearly with separation, 
and hence color charges are confined. How can one obtain a condensate 
of magnetic charges responsible for this? 
The idea is that due to a presumably educated gauge fixing 
the low-energy degrees of freedom of SU(N) Yang-Mills 
become transparent. Imposing a gauge condition 
invariant under the maximal Abelian subgroup 
U(1)$^{\tiny\mbox{N}-1}$, the emergence of chromo-magnetic monopoles 
can be observed. For example, demanding that the 
homogeneously transforming, hermitian field 
strength component $F_{23}(x)$ be 
diagonal after gauge fixing is no constraint 
for a gauge 
transformation$ \in $U(1)$^{\tiny\mbox{N}-1}$. 
Moreover, for a given configuration $A_\mu$ 
there may be points in space 
where the 
eigenvalues of $F_{23}$ are degenerate. At such a location 
the entire gauge group is unconstrained by 
the gauge condition. With respect to {\sl Abelian} components 
of the gauge field $A_\mu$ one 
encounters radially 
directed magnetic flux with the magnetic 
field becoming singular at 
the singularity of the gauge fixing procedure and along a half-line 
starting there. A magnetic monopole together 
with its Dirac string is recovered. However, 
the statement that at the instant $x^0$ there sits a 
magnetic monopole at point $\vec x$ is a 
{\sl highly gauge variant} and arbitrary one. Instead of diagonalizing 
$F_{23}(x)$ under gauge transformations 
we could have chosen to diagonalize $F_{12}(x)$. Generically, 
this would have caused the gauge 
singularities to be 
distributed in a different way. This is a serious problem, 
and one only can hope that the dynamics itself 
proliferates the gauge {\sl invariant} existence 
of monopoles at low resolution. Another unanswered 
question arises in view of the monopole 
interpretation using Abelian 
components of the gauge fields $A_\mu$: 
What justifies the projection onto Abelian components? 
The only hint that this procedure 
describes the low-resolution physics properly 
comes from the lattice where the 
string tension indeed was found to be 
saturated by the projected dynamics and, even better, 
by the monopole dynamics alone \cite{Shiba,Yama,Cherno}. 
Analytical insight into this 
miracle may be obtained if the vacuum 
at low resolution can be shown to be governed by the dynamics of 
an adjoint Higgs model with the
gauge field being 
essentially pure gauge \cite{SurHof1}.  

\noindent For now let us be pragmatic. Point-like 
monopoles in space correspond to
line-like trajectories in spacetime. It can be shown that a summation over
these trajectories for monopole charge $|Q|=1$ in the partition function 
leads to a scalar 
field theory for the monopole sector \cite{Bard}. However, working with the
projected gauge field alone the 
theory necessarily contains 
non-local terms which one would like to avoid \cite{Bard}. 
The introduction of a twin set of 
gauge fields can cure this \cite{Bard,Zwanziger}. Projecting QCD, 
one obtains a theory of electrically and magnetically charged matter 
fields and sets of (N-1) electric and magnetic 
abelian gauge fields $\vec {A}_\mu$ and $\vec {B}_\mu$, respectively. 
A {\em phenomenological} 
self-interaction of the magnetic monopole fields 
is assumed to yield a spontaneously generated condensate breaking the 
U(1)$^{N-1}$ symmetry \cite{Suzuki}. As a result, the set of gauge fields $\vec {B}_\mu$, 
interacting with the magnetic charges 
\footnote{This sector is a dual Ginzburg-Landau theory.}, 
becomes massive, and a constant string-tension $\sigma$ emerges.

\section{The case of finite temperature}

It has been shown on the lattice (for example \cite{Ha}) that a 
thermalized SU(3) Yang-Mills theory undergoes a deconfinement phase 
transition at critical temperatures $T^{lat}_c\sim 200-300$ MeV. As explained above 
the starting point for an analytical 
investigation of this phenomenon is the dual 
Ginzburg-Landau theory given by the following 
Lagrangian \cite{Suzuki}
\eqb
\label{Lag}
{\cal L}_{DGL}=-\frac{1}{4}(\pd_\mu\vec{B}_\nu-\pd_\nu\vec{B}_\mu)^2+\sum_{k=1}^3\left\{\left|(i\pd_\mu-
g\vec{\ep}_k\cdot\vec{B}_\mu)\phi_k\right|^2-V_k(\phi_k,\bar{\phi}_k)\right\}\ .
\eqe
In contrast to the electric, Abelian gauge group the magnetic U(1)$^2$ 
(with gauge field $\vec{B}_\mu=(B^3_\mu,B^8_\mu)$) is believed to be 
spontaneously broken by nonvanishing VEVs of the monopole fields $\phi_k$.     
In Eq.\,(\ref{Lag}) $g\vec{\ep}_i$ denote the effective magnetic 
charges with $\vec{\ep}_1=(1,0),\ 
\vec{\ep}_2=(-1/2,-\sqrt{3}/2),\ \vec{\ep}_3=(-1/2,\sqrt{3}/2)$, and the fields $\phi_k$ 
satisfy the constraint $\sum_{k=1}^{3}\mbox{arg}\,\phi_k=0$ \cite{Suzuki}.

The potential $V\equiv \sum_{k=1}^3V_k(\phi_k,\bar{\phi}_k)$ 
is introduced {\em phenomenologically} to 
account for the self-interaction of the monopole 
fields $\phi_k$. 
In the framework of the dual Ginzburg-Landau theory the deconfinement phase transition 
at finite temperature has been first discussed 
in Ref.\,\cite{Monden}. In Ref.\,\cite{TokiT} 
the critical temperature $T_c$ of the deconfinement phase transition 
was determined as the point where a thermal 
one-loop effective potential, calculated in euclidean spacetime with compactified time dimension of size
$\beta\equiv 1/T$, starts possessing 
an absolute minimum with vanishing monopole VEV's. For the sake of renormalizability and the desired 
feature of spontaneous breaking of the magnetic gauge symmetry the potential $V$ for the 
monopole fields was chosen to be Higgs-like: $\lambda\sum_{k=1}^3\left(\bar{\phi}_k\phi_k-v^2\right)^2$.
 
\noindent However, since the Abelian description of gluodynamics can only be valid up to a certain 
resolution $\Lambda_b$\footnote{After all perturbative QCD works well for large momenta.} 
one may wonder whether a renormalizable 
monopole interaction is imperative. Introducing a pair of external, oppositely charged, 
static, electric color-charges and 
integrating out the electric and (massive) magnetic gauge bosons and the monopole 
fluctuations about their VEV's in quadratic approximation a critical 
tempertaure $T_c\sim 500$ MeV was obtained from the corresponding effective potential. 
This is too high. The result is not surprising since the effective potential was calculated by 
abusing perturbation theory ($\lambda\sim 25$ from a fit to the string tension $\sigma$!). 

\noindent Our approach is therefore different. We view the monopole 
potential to be a quantum effective potential
already. With the lattice-motivated assumption that the vacuum dynamics 
is {\sl dominated by the monopole fields} 
it is sufficient 
to look for solutions to the classical 
equations of motion of the monopole sector. 
The potential is then constructed such that\\  
\noindent {\bf 1)} it is gauge invariant under the magnetic U(1)$^2$,\\ 
\noindent {\bf 2)} it admits topologically non-trivial, BPS saturated 
solutions along the compact, euclidean time
dimension
\footnote{Recall, that in the
euclidean description finite temperature is implemented by compactifying the time dimension.}, and\\
\noindent {\bf 3)} it allows for {\sl one} topologically trivial, 
non-vanishing VEV for the monopole fields at zero temperature.\\ 
\noindent Thereby, the requirement of BPS saturation derives from the fact that 
we are interested in a vacuum description. The corresponding 
fields must then saturate the 
lowest bound for the euclidean action. From {\bf 1)} and {\bf 2)} it follows that 
the ``square root'' $V_k^{1/2}$, defined as $V_k(\bar{\phi_k}\phi_k)\equiv V_k^{1/2}(\bar{\phi_k})
V_k^{1/2}(\phi_k)$, must have a single pole at $\phi_k=0$ \cite{ShifDva,Losev,Hof}. {\bf 3)} 
enforces an analytical part in order to obtain 
finite VEV's at zero temperature (stabilization). 
Furthermore, this analytical part can only be a power $\sim \phi^N$ 
since a genuine polynomial would introduce non-degenerate vacua. From 
{\bf 1)} it finally follows that 
\eqb
\label{Veff}
V_k(\phi_k,\bar{\phi}_k)=
=\lim_{N\to\infty}\,\left\{\frac{\Lambda^6}{\bar{\phi}_k\phi_k}+\kappa^2\Lambda^{-2(N-2)}
(\bar{\phi}_k\phi_k)^N-2\,\kappa\Lambda^{5-N}\frac{1}{\bar{\phi}_k\phi_k}\mbox{Re}\,\phi_k^{N+1}\right\}\ . 
\eqe
Thereby, $\Lambda$ is a mass-parameter, and $\kappa$ is 
some dimensionless coupling constant. Considering the rhs of Eq.\,(\ref{Veff}) at finite $N$, the 
potential explicitely breaks the magnetic U(1)$^2$ 
gauge symmetry, $\vec{B}_\mu\,\to\, \vec{B}_\mu+1/g\,\pd_\mu\vec{\theta}(x)\ ,\ \ \ 
\phi_k\,\to\,\e^{-i\vec{\ep}_k\cdot\vec{\theta}(x)}\phi_k$, 
down to Z$^2_{N+1}$ due to the term $\mbox{Re}\,\phi_k^{N+1}$ \cite{Hof}. 
Only in the limit of large $N$ is the gauge symmetry restored. 
For the purpose of qualitative illustration Fig. 1 shows the potential $V_k$. 
\begin{figure}
\vspace{5cm}
\includegraphics{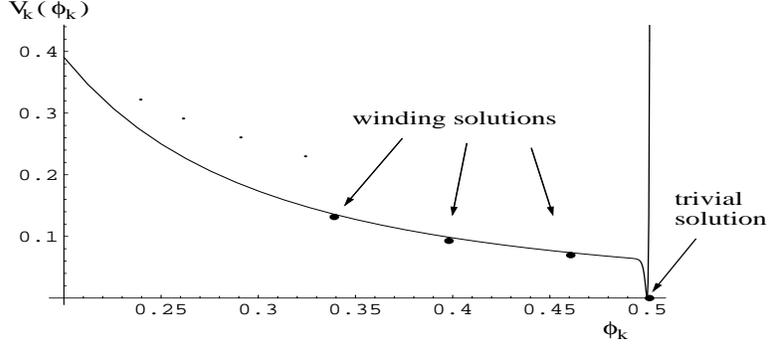}
\caption{The monopole potential $V_k$ as a 
function of real $\phi_k$ for $\kappa=1,\,\Lambda=0.5$, and $N=500$. Indicated are 
the potential energy densities of topologically trivial and 
non-trivial solutions to the BPS equations.} 
\label{} 
\end{figure}
Note that at $T=0$ this effective potential $V$ ideed 
does not allow for fluctuations of the fields $\phi_k$ due to its infinite
curvature at $\phi_k=\Lambda$. The BPS equations, corresponding to 
the potential of Eq.\,(\ref{Veff}), read
\eqb
\label{BPS}
\pd_\tau\phi_k=\bar{V}^k_{1/2}(\bar{\phi_k})\ ,\ \ \pd_\tau\bar{\phi}_k=V^k_{1/2}(\phi_k)\ .
\eqe
The rhs are only fixed up to phase factors $\e^{i\delta}$, $\e^{-i\delta}$, 
respectively. From Eq.\,(\ref{Veff}) we have for {\em periodic} solutions 
\eqb
\label{V1/2}
V^k_{1/2}(\phi_k)=\left\{\begin{array}{c} \pm i\frac{\Lambda^3}{\phi_k}\ ,\ \ \ \  (|\phi_k|<\Lambda)\\ 
0\ ,\ \ \ \ \ \ \ \ \ (|\phi_k|=\Lambda)\\ 
\ \infty\ ,\ \ \ \ \ \ \ (|\phi_k|>\Lambda)\end{array}\right.\ .
\eqe

\section{BPS saturated winding solutions}

For the case $|\phi_k|<\Lambda$ periodic solutions ($\phi_k(0)=\phi_k(\beta)$) 
to Eqs.\,(\ref{BPS}) subject to Eq.\,(\ref{V1/2}) have been discussed 
in Refs.\,\cite{ShifDva,Losev} within the framework of supersymmetric theories. For a compact, 
euclidean time dimension of length $\beta\equiv 1/T$ 
the set of topologically distinct solutions is
\eqb
\label{wind}
\phi^{n_k}_k(\tau)=\sqrt{\frac{\Lambda^3\beta}{2|n_k|\pi}}\,
\e^{2n_k\pi i\frac{\tau}{\beta}}\ ,\ \ (n_k\in{\bf Z})\ .
\eqe
Thereby, the sign of $n_k$ corresponds to the choice of 
phase in the BPS equations. 
Due to the phase constraint the sets of 
solutions $(\phi^{n_1}_1,\phi^{n_2}_2,\phi^{n_3}_3)$ can be labeled by 
the integers $n_1$ and 
$m$, which are both odd {\em or} even. The winding numbers 
$n_2,\,n_3$ are then given as $-n_1/2\mp m/2$, respectively. 
The gauge function 
$\vec{\theta}_{n_1,m}(\tau)$, transforming to the unitary gauge 
Re\,$\phi_k(\tau)>0$, Im\,$\phi_k(\tau)=0$, reads
\eqb
\label{uni}
\vec{\theta}_{n_1,m}(\tau)=\frac{2\pi}{\beta}\,
\tau\left(\begin{array}{c}n_1\\ \frac{m}{\sqrt{3}}\end{array}\right)\ .
\eqe
Note that this non-periodic function leaves the periodicity 
of the gauge field $\vec{B}_\mu$ intact.

\section{Deconfinement phase transition}

Eqs.\,(\ref{BPS}) admit the solution $\phi^0_k\equiv\Lambda$ with winding number 
and energy density zero, and the winding solutions of the previous section. 
We identify the set $(\phi^0_1,\phi^0_2,\phi^0_3)$ 
with the confining vacuum at low temperatures. In this regime 
the spectrum consists of glueballs with masses that are 
much larger than the prevailing 
temperatures \cite{Toki0}. The thermal equilibrium between the 
vacuum "medium" and its excitations is essentially 
realized at pressure zero, which indeed is 
observed on the lattice \cite{Beinl}. Above the deconfinement 
phase transition the ground state "medium" readily emits and absorbs 
(almost) free gluons under the 
influence of the heat bath like a black body emits and 
absorbs photons. We identify the genuine winding set 
of lowest potential energy density\footnote{By "genuine" we mean that 
each of the solutions $\phi_k$ is winding. 
Note, however, that 
there are also semi-winding sets, for example $(n_1,m)=(1,1)$, 
which contain two winding fields 
and one field of winding number zero. Since none of the 
monopole fields $\phi_k$ should be singled out it is natural
to assume that either all $\phi_k$ are winding or none at all.}, 
represented by 
$(n_1,m)=(2,0)$, with the ground 
state just above the deconfinement transition. 
In this picture the confining vacuum is a perfect thermal insulator 
up to the transition, 
where its structure drastically changes by the absorption 
of an amount of latent heat per volume \cite{Beinl} 
equal to the gap $\Delta\ep^{(2,0)}$ between the potential $V$ of the 
zero winding and the lowest genuine winding set. 
We estimate the critical temperature $T_c$ 
of this transition by assuming 
the deconfining vacuum to behave like an 
incompressible, static fluid with traceless energy-momentum-tensor \cite{Cley} 
(no scale anomaly) which is in thermal equilibrium with an ideal gas of gluons. In this case we 
have the equation of state $\ep=3p$ for pressure $p$ and energy density $\ep$ 
for both the vacuum (vac) and the gluon gas (gg). The equilibrium 
condition of equal pressures, $p_{vac}=p_{gg}$, 
then takes the following form \cite{Cley}
\eqb
\label{eq} 
\Delta\ep^{(2,0)}=\frac{8}{15}\pi^2 T_c^4\ ,\ \ \mbox{where}\ \ \ 
\Delta\ep^{(2,0)}=8\pi\Lambda^3T_c\ .
\eqe
Using Eqs.\,(\ref{eq}) and the condition $p_{vac}=p_{gg}$, we obtain $T_c=
\left(\frac{15}{\pi}\right)^{\frac{1}{3}}\Lambda\sim 1.68\,\Lambda$. 
Adopting the value $\Lambda=0.126$ GeV for the monopole 
condensate at $T=0$ from Refs.\,\cite{Toki0,TokiT}, this yields 
$T_c=0.212$ GeV which is compatible with the 
lattice results of $T_c=0.2-0.3$ GeV.

\section{The deconfining phase}

Transforming the winding solution of Eq.\,(\ref{wind}) to 
the unitary gauge by means of Eq.\,(\ref{uni}), 
we obtain the following expression 
for the individual condensate, corresponding to winding $n_k$:  
$\phi_k^{n_k}=\frac{\Lambda^{3/2}}{\sqrt{2\pi |n_k| T_l}}=\frac{\Lambda}{\sqrt{(30\pi^2
l)^{1/3}|n_k|}}$\footnote{$l\equiv\sum_k|n_k|\ .$}. 
For the configuration $(n_1,m)=(2,0)$ we obtain $\phi^{(2,0)}_{av}\sim 0.28\,\Lambda$. 
In Fig. 2 the evolution of the average monopole 
condensate is depicted up to $T\sim0.34$ GeV. 
\begin{figure}
\vspace{5.0cm}
\includegraphics{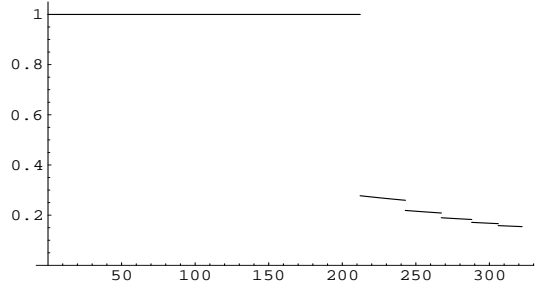}
\caption{The average monopole field $\phi_{av}(T)$ in units of 
$\phi_{av}(T=0)$. For further explanation see text.} 
\label{} 
\end{figure}
In unitary gauge the monopole fields are real, and 
the spontaneous breakdown of the U(1)$^2$ magnetic gauge symmetry then transparently 
generates a mass $m_B$ for the field $\vec{B}_\mu$ which scales linearly 
with the average monopole condensate, $m_B^{(n_1,m)}=\sqrt{3}\,g\,\phi^{(n_1,m)}_{av}$. 
In Refs.\,\cite{Toki0,TokiT} the value $m_B\sim 0.5$ GeV was 
determined at $T=0$, yielding a magnetic gauge coupling of $g=2.3$.
   
\noindent How does the string tension $\sigma$ evolve? An analytical 
expression for $\sigma$ in terms of $g,\,m_B$, and $m_\phi$ 
was derived in Ref.\,\cite{Toki0} by calculating the potential between 
two heavy and static quarks of opposite color charge, separated by a distance $R$. 
Thereby, an effective Lagrangian 
for the dynamics of the electric $U(1)^2$ gauge field $\vec{A}_\mu$ 
is obtained by integrating out the massive, magnetic vector field 
$\vec{B}_\mu$ and by assuming that the monopole fields do 
not fluctuate about their VEVs. Applying convenient gauge fixing terms and 
introducing external, static color currents, which couple to $\vec{A}_0$, one obtains an 
effective action. From this action one extracts the potential energy $V_{\bar{q}q}(R)$ and reads off as 
$\sigma=2\pi m_B^2/(3g^2)\ln\left[1+m_\phi^2/m_B^2\right]$. Since {\em static} test charges 
were assumed finite temperature does not necessitate any new considerations. The appearance of the 
monopole mass $m_\phi>m_B$ under the logarithm in the expression for $\sigma$ is due to its role as an ultraviolet 
cutoff for the integration over transverse momenta in the expression for 
the linear part of the potential $V_{\bar{q}q}(R)$. 
This is justified by the observation that $m_B\sim0$ inside the 
flux tube of radius $\rho\sim m_\phi^{-1}$ in a type II dual superconductor 
(see Ref.\,\cite{TokiT} and Refs. therein). Since we work with effective fields $\phi_k$ we cannot 
asign a mass to the fluctuations of the monopole fields about the 
background of a classical ground state. We may, however, assume that $\rho^{-1}$ 
scales with temperature in the same way as $m_B$ does. Since the dependence of $\sigma$ 
on the cutoff is logarithmic the result is not dramatically 
sensitive to the exact $T$ dependence of 
$\rho^{-1}$, which we will make explicit by distinguishing the following cases:\\  
(a) $\rho^{-1}(0)=\rho^{-1}(T)=1.26$ GeV \cite{Toki0} $\Rightarrow$ 
\eqb
\frac{\sigma(T_c)}{\sigma(0)}\sim (0.28)^2\frac{\ln(1+(1.26)^2/(0.5\times0.28)^2)}
{\ln(1+(1.26)^2/(0.5)^2)}\sim 0.19\ .
\eqe
(b) $\rho^{-1}(T)\propto m_B(T)$ $\Rightarrow$ 
\eqb
\frac{\sigma(T_c)}{\sigma(0)}\sim (0.28)^2\sim 0.08\ .
\eqe
Hence, the discrepency amounts to a factor 2 for these two extreme cases.

\section{Summary and discussion}

We considered a thermalized dual Ginzburg-Landau theory modelling hot SU(3) Yang-Mills theory. 
From the postulate that in contrast to the confining phase 
the ground state of the deconfining phase is characterized by 
topologically nontrivial, BPS saturated solutions to the 
classical equations of motion of the monopole sector and the uniqueness of the vacuum at $T=0$ 
we deviced the corresponding, 
gauge invariant interaction. As a consequence, the average monopole 
field undergoes a drastic decrease to about 1/4 of its 
zero temperature value across the phase boundary. Scaling linearly 
with the monopole condensate, 
the same applies to the mass of the magnetic vector fields generated by the 
spontaneous breaking of the magnetic U(1)$^2$ gauge symmetry. Working with the zero 
temperature value for the monopole condensate of Refs.\,\cite{Toki0,TokiT}, 
we obtain a reasonable critical temperature $T_c\sim 212$ MeV in 
contrast to the effective potential calculation 
of Ref.\,\cite{TokiT}, where $T_c\sim 500$ MeV, 
and the more realistic value of $T_c\sim200$ MeV was obtained by 
introducing an ad hoc $T$-dependence of the dimensionless 
coupling constant $\lambda$. 

\noindent Depending on the assumption about 
the $T$-dependence of the flux tube radius $\rho$ the string tension $\sigma$ decreases 
to about 1/12..1/5 of its value at $T=0$ across the phase boundary which is 
compatible with lattice measurements \cite{Beinl}. To determine $T_c$ 
we assumed a {\em free} gluon gas although close to the critical $T_c$ the lattice indicates 
a pressure in the deconfining phase which is sizably smaller than the
Stefan-Boltzmann asymptotics. However, since the leading term in an expansion of $p$ 
in (mass scale)/$T$ is quartic in $T$ we expect $T_c$ to be 
robust against changes in the subleading terms\footnote{If there were no subleading terms $T_c$ would be 
given by a cubic root.}. The presence of non-perturbative effects is described 
by non-vanishing values of the monopole condensate 
$\phi_{av}$ and the string tension 
$\sigma$, and the model predicts a slow decrease of these quantities. 
On the lattice and in effective potential calculations the deconfinement 
phase transition has been determined 
to be of first order \cite{TokiT,Beinl,Karsch}. In contrast, SU(2) Yang-Mills theory 
exhibits a second order phase transition \cite{Karsch}. The 
model of the present work can be easily adapted to 
the dual Ginzburg-Landau  
theory describing maximal abelian 
gauge fixed SU(2) Yang-Mills theory. The qualitative results 
would then be the same. It is clear that we cannot resolve this difference 
since by definition the order parameter $|\phi_{av}|$ is 
always discontinuous. However, the model gives a reason for the residual, non-trivial structure 
of monopole condensation in SU(2) and SU(3) Yang-Mills theories at high temperatures, it predicts a 
reasonable value for the critical temperature $T_c$, and it 
incorporates the low-temperature limit in the form 
of a constant solution to the BPS equations. The existence of a non-trivial structure 
of monopole condensation above the phase transition has been pointed out in Ref.\,\cite{Lau} 
a long time ago based on lattice investigations.

\section*{Acknowledgments}

The author would like to thank the organizers 
of ``Lepton Scattering, Hadrons and QCD'' 
for a very interesting workshop and a stimulating atmosphere.

\end{document}